\documentclass[conference]{IEEEtran}
\IEEEoverridecommandlockouts
\usepackage{cite}
\usepackage{amsmath,amssymb,amsfonts}
\usepackage{algorithmic}
\usepackage{graphicx}
\usepackage{textcomp}
\usepackage{xcolor}
\usepackage{tikz}
\usepackage{float}
\usepackage{comment}
\usetikzlibrary{shapes.geometric, positioning, arrows, calc, arrows.meta}

\def\BibTeX{{\rm B\kern-.05em{\sc i\kern-.025em b}\kern-.08em
    T\kern-.1667em\lower.7ex\hbox{E}\kern-.125emX}}

\def\BibTeX{{\rm B\kern-.05em{\sc i\kern-.025em b}\kern-.08em
    T\kern-.1667em\lower.7ex\hbox{E}\kern-.125emX}}
\begin{document}

\title{Simulation-Driven Evaluation of Chiplet-Based Architectures Using VisualSim\\
}

\author{\IEEEauthorblockN{Wajid Ali}
\IEEEauthorblockA{\textit{University of Engineering and Technology} \\
\textit{}
Lahore, Pakistan \\
2021ee79@student.uet.edu.pk}
\and
\IEEEauthorblockN{Ayaz Akram}
\IEEEauthorblockA{\textit{Samsung} \\
San Jose, USA \\
ayaz.akram@samsung.com}
\and
\IEEEauthorblockN{Deepak Shankar}
\IEEEauthorblockA{\textit{Mirabilis Design Inc} \\
Santa Clara, USA \\
deepak.shankar@mirabilisdesign.com}
}

\maketitle
\begin{abstract}

%This paper looks at the design, modeling, and simulation of a multi-die System-on-Chip (SoC) architecture using VisualSim. It focuses on chiplet technology, which offers a new solution to problems in modern chip designs. Traditional monolithic chips are becoming harder to manufacture due to high costs, power issues, and limits in performance scaling. Chiplets, small modular silicon units, provide a flexible, scalable, and efficient way to solve these challenges by combining them into a single package. In this study, we created and tested a system with multiple chiplets. Each chiplet has multicore processor clusters connected by a CMN600 network-on-chip (NoC) for smooth internal communication [9][10]. The architecture can grow by adding more chiplets and cores to handle increasing demands. Using VisualSim, we analyzed the system’s performance under different workloads, focusing on communication delays, memory usage, workload distribution, and power vs. performance balance. These results help in improving chiplet-based designs and show their potential for building cost-effective and scalable semiconductor systems.

This paper focuses on the simulation of multi-die System-on-Chip (SoC) architectures using VisualSim, emphasizing chiplet-based system modeling and performance analysis. Chiplet technology presents a promising alternative to traditional monolithic chips, which face increasing challenges in manufacturing costs, power efficiency, and performance scaling. By integrating multiple small modular silicon units into a single package, chiplet-based architectures offer greater flexibility and scalability at a lower overall cost. In this study, we developed a detailed simulation model of a chiplet-based system, incorporating multicore ARM processor clusters interconnected through a ARM CMN600 network-on-chip (NoC) for efficient communication~\cite{friese2024exploring,pellegrini2020arm}. The simulation framework in VisualSim enables the evaluation of critical system metrics, including inter-chiplet communication latency, memory access efficiency, workload distribution, and the power-performance tradeoff under various workloads. Through simulation-driven insights, this research highlights key factors influencing chiplet system performance and provides a foundation for optimizing future chiplet-based semiconductor designs.
\end{abstract}
\section{Introduction}
The semiconductor industry is facing significant challenges as traditional single-chip designs struggle to meet increasing demands for power, efficiency, and scalability. Moore's Law, which predicted regular improvements in chip performance, is slowing, and Dennard Scaling, which helped improve performance by shrinking transistors, is no longer effective. Additionally, manufacturing costs are rising, and adding more features to a single chip has become increasingly complex. These trends signal the end of Moore’s Law and mark the beginning of new approaches in computing, such as the introduction of advanced architectures and technologies like chiplets~\cite{theis2017end}.\par
Chiplets are small pieces of silicon that can be combined in one package to make a bigger and more flexible System-on-Chip (SoC). Instead of using one big chip, chiplet-based designs use smaller parts that are made for specific tasks. These parts communicate with each other using advanced connection methods like the Universal Chiplet Interconnect Express (UCIe)~\cite{sharma2022universal}. This approach offers several benefits, such as reduced manufacturing costs, improved yields, and the ability to design chips tailored for specific applications~\cite{li2020chiplet,muhammad2010virtual,chidella2018novel}. Chiplets also provide scalability, as additional units can be integrated to handle more tasks without redesigning the entire system. However, designing and optimizing chiplet-based systems presents challenges. Engineers must consider memory operations, inter-chiplet communication, power usage, and thermal management to achieve efficient performance~\cite{li2020chiplet,muhammad2010virtual}.\par
To handle these challenges, tools like VisualSim Architect are very important. VisualSim Architect is a platform that helps engineers model and test complex chip designs in a virtual environment. It allows engineers to create detailed models of chip components, simulate how the system will behave under different conditions, and study important results like speed, power consumption, data transfer efficiency, and heat generation. By using VisualSim, engineers can identify potential problems, optimize performance, and make better decisions during the design phase.\par
VisualSim Architect has a user-friendly graphical interface, making it accessible for both beginners and experts. It also includes a large library of ready-made components, such as processors, memory units, and interconnects, which can be easily combined to build complex Systems-on-Chip (SoC) architectures. For chiplet-based designs, VisualSim allows engineers to study how the chiplets communicate, analyze memory access patterns, and explore different interconnect settings to ensure the system performs efficiently~\cite{muhammad2010virtual,chidella2018novel,asaduzzaman2007performance}. \par
One of the most important features of VisualSim is its ability to simulate real-world and dynamic workloads. This helps engineers see how their designs will perform under actual usage and different conditions. For example, they can test what happens when more cores are added to a chiplet, analyze memory access patterns, or check how changes in interconnect settings like bandwidth and latency affect performance. These simulations give engineers valuable insights, helping them identify bottlenecks, optimize designs, and make better decisions before manufacturing. This reduces the need for expensive prototypes, saves time, lowers development costs, and ensures efficient, scalable, and reliable chip designs for modern applications~\cite{muhammad2010virtual,asaduzzaman2007performance}. \par
In this study, we use VisualSim Architect to design, model, and test a chiplet-based multi-die System-on-Chip (SoC). Each chiplet contains processor clusters that communicate within the chiplet using a CMN600 network-on-chip (NoC)~\cite{bjerregaard2006survey}. The chiplets are connected to each other using UCIe links, which ensure smooth communication between them. Our study focuses on three main goals: (1) studying how different memory setups, like shared DRAM in a single-chip design and separate DRAM modules in a chiplet-based design, affect performance and scalability, (2) examining how adding more cores in chiplets impacts processing power and the balance between performance and energy use, and (3) testing how interconnect technologies, such as UCIe and CMN600~\cite{sharma2022universal,pellegrini2020arm}, help chiplets communicate and transfer data efficiently in scalable systems.\par
\section{Methodology}
This study evaluates a multi-die System-on-Chip (SoC) architecture using VisualSim Architect, a simulation platform designed for detailed modeling and performance analysis. The focus is on understanding how various architectural configurations, including monolithic and chiplet-based designs, perform under real-world workloads. The experiments explore the impact of factors such as memory hierarchy, core distribution, and interconnect technologies on key performance metrics like latency, power consumption, and scalability. By leveraging VisualSim Architect, the study models complex components like CPUs, caches, memory controllers, and interconnects to simulate task execution and data flow within the system. This approach enables detailed analysis of how design changes affect system efficiency, helping optimize both monolithic and chiplet-based SoC designs. The methodology consists of two key phases: experimental setup and simulation framework to assess system performance under varying configurations.

\subsection{Experimental Setup}
The system is based on the ARM A720AE model, configured with multicore clusters, DRAM modules, and interconnections using CMN600 NoC and UCIe links. In Experiment 1, a single monolithic chip with two ARM A720AE clusters is used. Each cluster contains 1 core with private L1 Data and Instruction Caches, a private L2 Cache, and a shared L3 Cache. Both clusters are connected via a CMN600 Coherent Mesh Network (NoC), and a two DRAM module is used for memory requests from both clusters. This setup allows for scalability to add more cores. \par
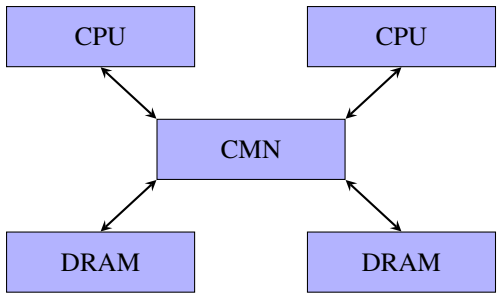
\begin{figure}[h!]
    \centering
    \begin{tikzpicture}[
        block/.style={rectangle, draw, fill=blue!30, text centered, minimum width=2.5cm, minimum height=0.8cm},
        arrow/.style={draw, thick, <->, >=stealth},
        doublearrow/.style={draw, thick, <->, >=stealth, double distance=2pt}
    ]
    % CPU blocks
    \node[block] (cpu1) at (0,0) {CPU};
    \node[block] (cpu2) at (4,0) {CPU};

    % CMN block
    \node[block] (cmn) at (2,-1.5) {CMN};

    % DRAM blocks
    \node[block] (dram1) at (0,-3) {DRAM};
    \node[block] (dram2) at (4,-3) {DRAM};

    % Double Arrows
    \draw[arrow] (cpu1.south) -- (cmn.north west);
    \draw[arrow] (cpu2.south) -- (cmn.north east);
    \draw[arrow] (dram1.north) -- (cmn.south west);
    \draw[arrow] (dram2.north) -- (cmn.south east);

    \end{tikzpicture}
    \caption{Block diagram of Experiment 1: Monolithic SoC}
    \label{fig:exp1}
\end{figure}
In Experiment 2, the system is expanded to a chiplet-based design with two ARM A720AE clusters placed on separate dies. The clusters are connected using UCIe interconnects, and each chiplet retains its private cache hierarchy and a dedicated DRAM module. All other parameters remain the same as in Experiment 1. 

\begin{figure}[H]
    \centering
    \begin{tikzpicture}[
        block/.style={rectangle, draw, fill=blue!30, text centered, minimum width=2.5cm, minimum height=0.8cm},
        arrow/.style={draw, thick, <->, >=Latex}
    ]
    % CPU blocks
    \node[block] (cpu1) at (0, 0) {CPU};
    \node[block] (cpu2) at (4, 0) {CPU};

    % CMN blocks
    \node[block] (cmn1) at (0, -1.8) {CMN};
    \node[block] (cmn2) at (4, -1.8) {CMN};

    % DRAM blocks
    \node[block] (dram1) at (0, -3.6) {DRAM};
    \node[block] (dram2) at (4, -3.6) {DRAM};

    % Arrows and Label
    \draw[arrow] (cmn1.east) -- (cmn2.west) node[midway, above] {UCIe CHI};
    \draw[arrow] (cmn1.north) -- (cpu1.south);
    \draw[arrow] (cmn2.north) -- (cpu2.south);
    \draw[arrow] (dram1.north) -- (cmn1.south);
    \draw[arrow] (dram2.north) -- (cmn2.south);
    \end{tikzpicture}
    \caption{Block diagram of Experiment 2: Chiplet-Based SoC.}
    \label{fig:exp2}
\end{figure}
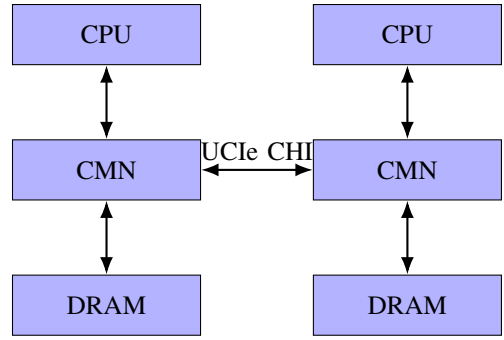

In Experiment 3, the system retains the chiplet-based design but increases the number of cluster from 2 to 4, significantly enhancing computational capacity. Each chiplet is equipped with two DRAM modules connected via CMN600, providing additional memory bandwidth.

\begin{figure}[h!]
    \centering
    \begin{tikzpicture}[
        block/.style={rectangle, draw, fill=blue!30, text centered, minimum width=2cm, minimum height=0.8cm},
        arrow/.style={draw, thick, <->, >=stealth}
    ]
    % Top Layer
    \node[block] (dram1_top) at (0, 0) {DRAM};
    \node[block] (cmn1_top) at (0, -1.5) {CMN};
    \node[block] (cpu1_top) at (-2.5, -1.5) {CPU};
    \node[block] (cmn2_top) at (2.5, -1.5) {CMN};
    \node[block] (dram2_top) at (2.5, 0) {DRAM};
    \node[block] (cpu2_top) at (5, -1.5) {CPU};

    % Bottom Layer
    \node[block] (dram1_bottom) at (0, -4.5) {DRAM};
    \node[block] (cmn1_bottom) at (0, -3) {CMN};
    \node[block] (cpu1_bottom) at (-2.5, -3) {CPU};
    \node[block] (cmn2_bottom) at (2.5, -3) {CMN};
    \node[block] (dram2_bottom) at (2.5, -4.5) {DRAM};
    \node[block] (cpu2_bottom) at (5, -3) {CPU};

    % Arrows and Labels (Top Layer)
    \draw[arrow] (cmn1_top.east) -- (cmn2_top.west) node[midway, above, xshift=-0.08cm, yshift=0.3cm] {UCIe CHI};
    \draw[arrow] (cmn1_top.north) -- (dram1_top.south);
    \draw[arrow] (cmn2_top.north) -- (dram2_top.south);
    \draw[arrow] (cpu1_top.east) -- (cmn1_top.west);
    \draw[arrow] (cmn2_top.east) -- (cpu2_top.west);

    % Arrows and Labels (Bottom Layer)
    \draw[arrow] (cmn1_bottom.east) -- (cmn2_bottom.west) node[midway, above, xshift=-0.08cm, yshift=0.3cm] {UCIe CHI};
    \draw[arrow] (dram1_bottom.north) -- (cmn1_bottom.south);
    \draw[arrow] (dram2_bottom.north) -- (cmn2_bottom.south);
    \draw[arrow] (cpu1_bottom.east) -- (cmn1_bottom.west);
    \draw[arrow] (cmn2_bottom.east) -- (cpu2_bottom.west);

    % Vertical Connection between Top and Bottom Layers
    \draw[arrow] (cmn1_top.south) -- (cmn1_bottom.north);
    \draw[arrow] (cmn2_top.south) -- (cmn2_bottom.north);
    \end{tikzpicture}
    \caption{Block diagram of Experiment 3: Multi-Chiplet SoC.}
    \label{fig:exp3}
\end{figure}
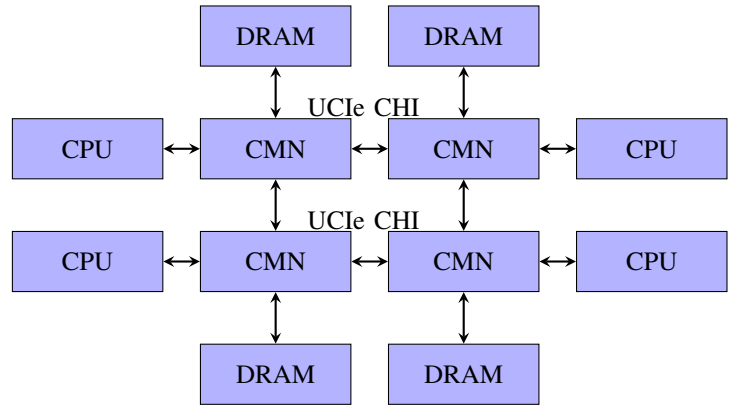

\subsection{Simulation Framework}
VisualSim Architect is used for modeling and simulation. This platform enables the creation of detailed models for system components, including task graphs, CPUs, caches, memory, and interconnects. The simulations analyze performance metrics such as latency, power consumption, and scalability under real-world workloads. Parameters like memory distribution settings, core cluster configurations, and UCIe link properties are adjusted to evaluate their impact on system behavior.

\section{Results and Discussions}
This study looks at the performance of micro-benchmark CCa on a multi-die System-on-Chip (SoC) by analyzing the results of the experiments. The main focus is on important performance factors like latency, power use, and scalability in different designs, including single-chip (monolithic) and chiplet-based systems. The study explains behavioral latency, power usage, and CMN600 interconnect latency in detail to show how changes in design and settings affect the system. These results help to understand the balance between performance and scalability and show how chiplet-based designs can be a good option for building scalable systems in the future.

\subsection{Behavioral Latency}
Behavioral latency shows how much time the CPU clusters take to finish a task. The Beh-Flow-Latency value tracks this time, helping us understand how well the system performs. It also helps find any delays or problems in how the CPU clusters handle tasks.
\begin{figure}[H]
    \centering
    \includegraphics[width=0.8\linewidth]{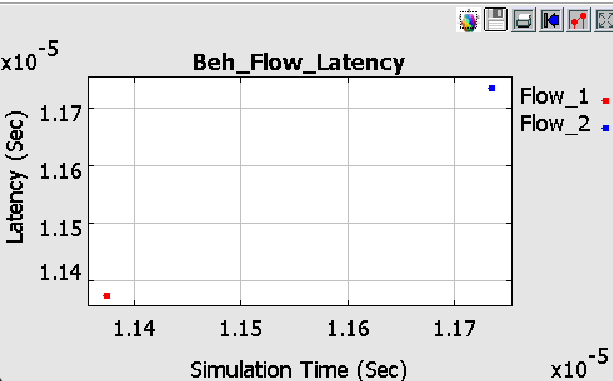} % Replace with your image file name
    \caption{Behavioral Latency: Task Completion Time by CPU Clusters Experiment 1}
    \label{fig:behavioral_latency}
\end{figure}
\begin{figure}[H]
    \centering
    \includegraphics[width=0.8\linewidth]{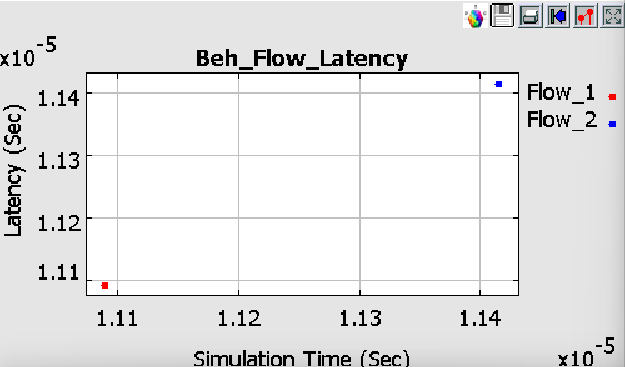} % Replace with your image file name
    \caption{Behavioral Latency: Task Completion Time by CPU Clusters Experiment 2}
    \label{fig:behavioral_latency}
\end{figure}
\begin{figure}[H]
    \centering
    \includegraphics[width=0.8\linewidth]{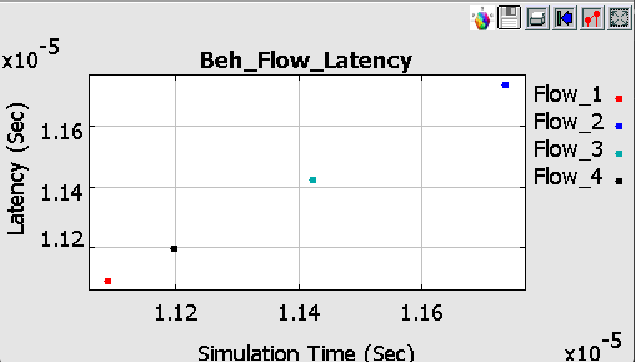} % Replace with your image file name
    \caption{Behavioral Latency: Task Completion Time by CPU Clusters Experiment 3}
    \label{fig:behavioral_latency}
\end{figure}
In Experiment 1, the monolithic chip design with two clusters completes the task in about $1.17 \times 10^{-5}$ seconds. In Experiment 2, the system changes to a chiplet-based design with two chiplets connected using UCIe interconnects. Here, the task completion time improves slightly to $1.14 \times 10^{-5}$ seconds
, thanks to the use of dedicated DRAM for each chiplet. In Experiment 3, the system is expanded to four clusters, increasing scalability in the chiplet-based design. The task completion time in this setup is around $1.16 \times 10^{-5}$ seconds
. While adding more clusters improves computational power, it also increases communication traffic between the chiplets, slightly increasing latency compared to Experiment 2. These results show the need to carefully balance cluster expansion and communication efficiency for better system performance.
\subsection{Power}
The power consumption analysis highlights how the different system configurations affect energy usage. It shows how factors like adding more cores, increasing clusters, and using dedicated DRAM modules influence the total power required. By comparing the experiments, we can see the trade-offs between higher power consumption and improved performance in different setups.
\begin{figure}[H]
    \centering
    \includegraphics[width=0.8\linewidth]{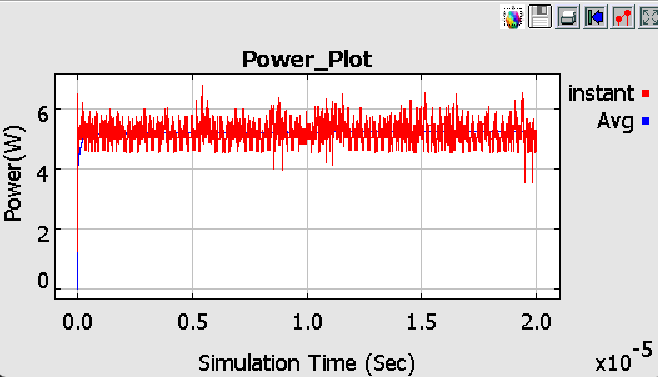}
    \caption{Power Plot: Instantaneous and Average Power over Simulation Time Exp 1}
    \label{fig:power_plot}
\end{figure}
In Experiment 1, the monolithic chip design with two clusters shows consistent power consumption because it uses a centralized memory and processing structure.This design keeps the power usage steady and predictable, as all memory requests are handled by the DRAM's on the same chip, with minimal overhead from inter-cluster communication.
\begin{figure}[H]
    \centering
    \includegraphics[width=0.8\linewidth]{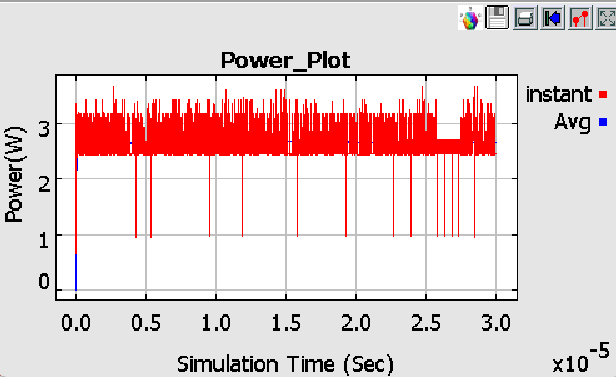}
    \caption{Power Plot: Instantaneous and Average Power over Simulation Time Exp 2 Die 1}
    \label{fig:power_plot}
\end{figure}
\begin{figure}[H]
    \centering
    \includegraphics[width=0.8\linewidth]{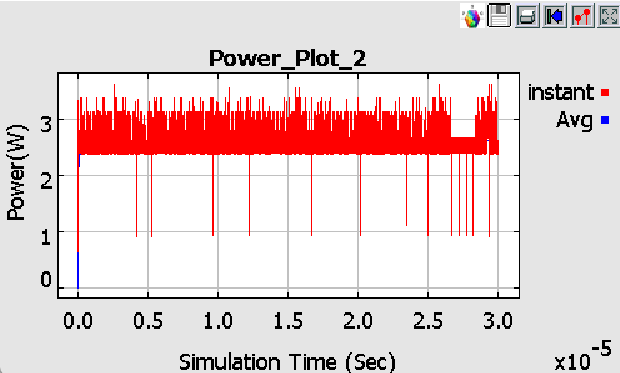}
    \caption{Power Plot: Instantaneous and Average Power over Simulation Time Exp 2 Die 2}
    \label{fig:power_plot}
\end{figure}
In Experiment 2, the shift to a chiplet-based design with two chiplets connected by UCIe interconnects increases power consumption slightly. This is because the UCIe interconnect adds a small overhead. However, the distributed memory architecture improves performance, making the power increase justifiable for scalability.
\begin{figure}[H]
    \centering
    \includegraphics[width=0.8\linewidth]{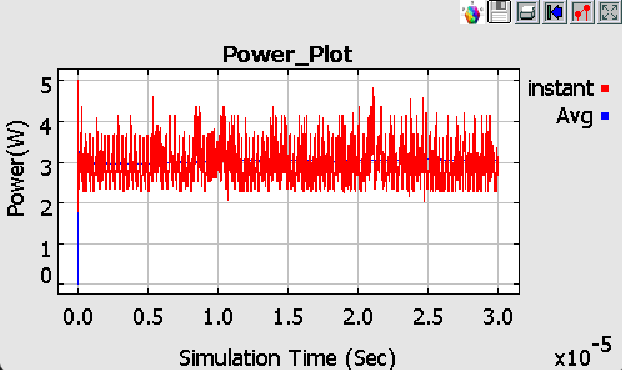}
    \caption{Power Plot: Instantaneous and Average Power over Simulation Time Exp 3 Die 1}
    \label{fig:power_plot}
\end{figure}
\begin{figure}[H]
    \centering
    \includegraphics[width=0.8\linewidth]{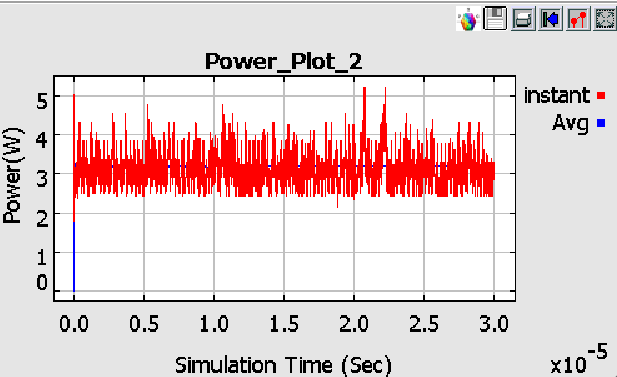}
    \caption{Power Plot: Instantaneous and Average Power over Simulation Time Exp 3 Die 2}
    \label{fig:power_plot}
\end{figure}
In Experiment 3, where the system expands from two clusters to four, power consumption increases further. This is due to the additional clusters and the two DRAM modules per chiplet. Although the system uses more power, the increased computational capacity and memory bandwidth support higher workloads, improving power efficiency per task completed. These results demonstrate the trade-offs between power consumption and performance as the system scales.

\section{Conclusion}
This study looked at the performance and scalability of a multi-die System-on-Chip (SoC) using VisualSim Architect. We compared monolithic and chiplet-based designs by analyzing key factors like latency and power consumption. The results showed that chiplet-based designs, with UCIe interconnects and separate DRAM modules, provide better scalability and performance than monolithic designs. However, they also use more power and have slightly higher communication delays.\par
Overall, chiplet-based designs are a promising solution for building scalable and efficient systems to meet modern computing needs. By carefully balancing memory, core scaling, and interconnects, these designs can handle increasing performance demands effectively. This study offers useful insights for improving future chiplets architectures.

\bibliographystyle{plain}
\bibliography{ref}

\begin{comment}

\end{comment}

\end{document}